# Can I Trust This Chatbot? Assessing User Privacy in AI-Healthcare Chatbot Applications

**Yener, Ramazan**       University of Illinois Urbana-Champaign, USA | ryener2@illinois.edu

**Chen, Guan-Hung**      University of Illinois Urbana-Champaign, USA | gc50@illinois.edu

**Gumusel, Ece**         Indiana University Bloomington, USA | egumusel@iu.edu

**Bashir, Masooda**      University of Illinois Urbana-Champaign, USA | mnb@illinois.edu

**ABSTRACT**

As Conversational Artificial Intelligence (AI) becomes more integrated into everyday life, AI-powered chatbot mobile applications are becoming increasingly adopted across industries, particularly in healthcare domain. These chatbots offer accessible and 24/7 support, yet their collection and processing of sensitive health data present critical privacy concerns. While prior research has examined chatbot security, privacy issues specific to AI healthcare chatbots have received limited attention. Our study evaluates the privacy practices of 12 widely downloaded AI healthcare chatbot apps available on the App Store and Google Play in the United States. We conducted a three-step assessment analyzing: (1) privacy settings during sign-up, (2) in-app privacy controls, and (3) the content of privacy policies. The analysis identified significant gaps in user data protection. Our findings reveal that half of the examined apps did not present a privacy policy during sign up, and only two provided an option to disable data sharing at that stage. The majority of apps' privacy policies failed to address data protection measures. Moreover, users had minimal control over their personal data. The study provides key insights for information science researchers, developers and policymakers to improve privacy protections in AI healthcare chatbot apps.

**INTRODUCTION**

Artificial Intelligence (AI) has swiftly integrated into daily routines across diverse domains, simplifying tasks and offering on-demand assistance for everything from online interactions to personal health decisions. This rapid progress has also prompted society to confront both the potential advantages and the possible pitfalls of an AI-driven world. Although AI solutions can significantly enhance accessibility, efficiency, and overall quality of life, they raise fundamental questions around privacy, transparency, and accountability, particularly as people place growing levels of trust in AI systems to manage sensitive information. In healthcare, these issues are especially pressing, where AI chatbots promise around-the-clock guidance and personalized support but at the same time collect large amounts of personal data that must be protected. Recognizing both the promise and the risks of AI, it is crucial for individuals, policymakers, technology developers and researchers to responsibly address its far-reaching social impact, ensuring that trust in these tools can be built on robust safeguards and ethical design.

Within the broader field of Information Science which encompasses areas such as knowledge organization, digital curation, and user experience, AI stands as a transformative force reshaping how data and knowledge are produced, disseminated, and safeguarded. As AI-driven tools continue to expand in both scope and sophistication, new challenges and opportunities surface around data ethics, algorithmic accountability, and the equitable provision of informational services. By addressing issues such as privacy, transparency, and user empowerment, the information science community can ensure that AI's rapid growth is guided by ethical frameworks and robust governance structures.

Conversational AI systems, powered by large language models (LLMs), now enable personalized, dynamic interactions across domains including healthcare. Chatbots like ADA (Jungmann et al., 2019), Woebot, Wysa, and Youper (Farzan et al., 2025) and MARVIN (Ma et al., 2025) have shown potential to support mental health and chronic condition management. However, due to the nature of AI, these healthcare chatbots collect and store personal data. Users may also be unaware of how their personal data is processed or they might not know that they have the right to control their data (Ischen et al., 2020; Ive et al., 2024). There is also lack of data transparency in these chatbots (Zhang et al., 2023). The existing literature focuses on technical aspects of data privacy and security issues including data breaches involving large language models (Bao et al., 2025; Surani & Das, 2022; Li, 2023; Yang et al., 2023). There remains a gap in research addressing end-users' privacy concerns in AI healthcare chatbots especially regarding data access, control, and deletion. To address this, we conducted a user-centered privacy evaluation of 12 widely used chatbot apps. Using a three-phase checklist, we assessed (1) how apps collect, process, and share personal data, (2) what controls users have to access, modify, or delete their data, and (3) how transparent and compliant the apps are with privacy regulations including The General Data Protection Regulation (GDPR), the Health Insurance Portability and Accountability Act (HIPAA), and The California Consumer Privacy Act (CCPA). Our findings reveal uneven disclosures, limited user control, and inconsistent compliance. By highlighting these issues, this study contributes a practical framework and user-focused insights to model the ethical development and regulation of AI-driven healthcare tools.

## BACKGROUND

### History of Mobile AI Healthcare Chatbot Apps

The first chatbot, ELIZA designed by Joseph Weizenbaum in 1966, was created to mimic natural conversations between humans and computers (Weizenbaum, 1966). ELIZA was revolutionary in demonstrating how computers could engage in structured dialogue with users, this laid the foundation for modern AI-powered virtual assistants. ELIZA showed AI-based systems could simulate human interaction and encourage users to express their thoughts, developing trust and engagement with computers.

Over the years, AI chatbots have advanced due to improvements in natural language processing (NLP) (Aslam, 2023), neural networks (Javed et al., 2023), and user-centered designs (Böhm et al., 2020), aiming to enhance the quality of interactions, improving efficiency and quality of customer services (Adamopoulou & Moussiades, 2020; Ekechi et al., 2024; Følstad et al., 2018; Gnewuch et al., 2017). The mobile revolution of the 2010s led to the emergence of new enhanced chatbots such as Siri, Alexa, and Google Assistant, transforming the way we communicate verbally and shaping speech patterns in various contexts (Djalolovna, 2024). With the rise of smartphones and mobile apps, many chatbots, including healthcare chatbots, have been adopted into individuals' daily lives, providing them with personalized and real-time support (Ciesla, 2024; Gamble, 2020). They now offer personalized medical advice, symptom checks, mental health counseling, and administrative help for both patients and healthcare professionals. A study by Laymouna et al. (2024) found that using chatbots in healthcare improves care quality by enhancing mental health support and self-management, increases accessibility for underserved populations, and reduces costs through administrative efficiency and scalable interventions.

Additionally, chatbots provide round-the-clock service for patients, are cost-effective, and offer information about symptoms, including mental health assistance, dermatology, and other health issues that individuals might face. Moreover, insufficient mental health providers have led individuals to seek alternative ways to access health facilities by using technology (Vaidyam et al., 2019). Recent findings indicate that many individuals now turn to AI chatbots for health advice. For example, Shahsavar & Choudhury (2023) revealed that most respondents (476 out of 607) were open to using ChatGPT for self-diagnosis. However, since ChatGPT is not specifically trained in medical literature, their study underscores the risk of relying on a general chatbot for clinical advice. These results support the need for more specialized, trustworthy healthcare chatbots, ones that incorporate domain-specific expertise and rigorous validation, so that users can safely and accurately address health concerns. Many existing chatbots are designed for a particular goals or specific user groups. For instance, chatbots such as Endurance support dementia patients, Casper aims to reduce loneliness in individuals with insomnia (Ayanouz et al., 2020) and Med-What answers basic health FAQs and provides information on diseases and symptoms (Pathan et al., 2021). Yet, many of these systems offer only "monotonous" or scripted question-and-answer interactions, limiting their ability to foster trust, capture nuanced patient inputs, or predict health conditions. While many AI healthcare chatbots are presented through web browser, they are now increasingly available on mobile apps, making them more accessible, engaging, and convenient for users. Although many AI healthcare chatbots are designed to provide helpful advice, they also raise important questions about how user data is handled.

### User Privacy Concerns in AI Healthcare Chatbot Apps

While mobile apps have significantly enhanced user engagement and expanded healthcare applications, their growing reliance on personal data processing has raised critical concerns regarding user privacy protection and regulatory compliance (Li, 2023). For instance, Sharma & Bashir (2020) examined COVID-19-related apps and found that these apps collect and store excessively sensitive data, including health data. In addition to excessive data collection and storage, privacy concerns related to user data in AI healthcare chatbot mobile apps which heavily rely on large datasets (Bouhia et al., 2022; Cai et al., 2023; Calvaresi et al., 2021) stem from various factors, including data misuse, lack of transparency, and inadequate user control. Studies have shown that users fear their personal data may be misused, leading to potential privacy violations and unauthorized access (Gumusel, 2024; Waheed et al., 2022). This concern is further compounded by declining trust in chatbots, as users become skeptical about how their sensitive information is handled (Belen Saglam et al., 2021). Moreover, many users are unaware or uninformed how their data is processed, primarily due to insufficient transparency and the absence of clear informed-consent mechanisms (Ive et al., 2024; Waheed et al., 2022). Without explicit disclosures about data collection, storage, and sharing practices, users are left with limited control over their information, increasing their vulnerability to privacy risks. Building trust requires not only robust technical safeguards but also clear, user-centered privacy policies that address these concerns. Factors like transparency in data handling, ease of access to privacy settings, and clear explanations of third-party data sharing play pivotal roles in fostering trust between users and chatbots. However, current studies insufficiently address user-perceived privacy concerns when interacting with third-party-provided mobile AI healthcare chatbot apps.

**Legal Compliance Challenges in AI Healthcare Chatbots**

Privacy regulations such as the GDPR, CCPA, and HIPAA play a crucial role in protecting individuals' personal data. HIPAA is a US federal law that protects patient privacy by regulating how healthcare organizations handle protected health information (PHI) (Rights (OCR), 2008). GDPR is Europe's primary privacy law that requires organizations to protect EU citizens' personal data (*Regulation - 2016/679 - EN - Gdpr - EUR-Lex*, n.d.). CCPA is also a privacy law that protects Californians' personal data by providing them with more control over their personal data (*California Consumer Privacy Act (CCPA)*, 2018). However, there is a notable gap when it comes to analyzing chatbots' privacy policies and their compliance with regulations from an organizational design perspective. For instance, a study by Li (2023) indicates that AI companies, such as OpenAI, do not comply with HIPAA, GDPR and CCPA. Thus, using AI in healthcare may pose risks to privacy and security. Additionally, research by Marks & Haupt (2023) state that HIPAA is outdated and no longer adequately protects individuals. For example, HIPAA does not cover digital platforms outside of healthcare organizations. If individuals utilize chatbots for health purposes outside of healthcare facilities, they are not subject to HIPAA regulations. Although regulatory frameworks such as GDPR, HIPAA, and CCPA aim to safeguard users' privacy, their application to conversational AI technologies remains limited because these laws were not specifically designed for AI systems; they govern general data privacy (Price & Cohen, 2019). Even though AI chatbots process vast amounts of personal data, HIPAA does not yet provide specific guidelines for compliance (Rezaeikhonakdar, 2023). Moreover, inconsistencies in enforcement across regions and a lack of AI-specific provisions in these regulations exacerbate the risks associated with sensitive healthcare data. This creates a pressing need for updated guidelines that account for the dynamic nature of AI-powered tools. Furthermore, research on AI chatbot privacy from users' perspective is relatively limited. Privacy concerns, such as transparency, consent, and trust in AI chatbots, are often overlooked, leaving users vulnerable to privacy risks without sufficient information (Hasal et al., 2021). However, there are only a few articles that delve into the design and improvement of the organization's privacy policy. Furthermore, not a lot of academic papers discussed this issue from a patient or customer perspective. Surani & Das (2022) argue that recent studies focus more on generating technological solutions to secure users' data rather than exploring the privacy aspects of the issues. Chametka et al. (2023) indicate that current studies mostly focus on certain aspects of security and privacy, or the experience of non-adopters of AI-based chatbots. Consequently, by downloading and using AI healthcare chatbots, we intend to explore how they handle privacy from the perspective of actual users. This will help improve privacy policy designs, ensuring they cater to both technical and user-focused needs.

**METHODOLOGY**

We employed a multi-step approach, which included the selection of chatbot applications, analyzing privacy practices during sign-up, a user-centered evaluation of application experiences, and an in-depth review of privacy policies. To support this evaluation, we developed a three-phase Privacy Evaluation Checklist designed to systematically assess the privacy practices of these chatbots. Our methodology evaluates not only the content of privacy policies but also their accessibility and usability from the user's perspective, aiming to highlight potential privacy risks.

**App Selection**

The app selection was limited to those available in the United States. Both the Apple App Store and Google Play Store have categories labeled "Health and Fitness" and "Medical." Initially, we checked these categories to identify top-ranking apps related to health and medical use. However, we found that we could not search within these categories, making it difficult to locate AI healthcare chatbots directly. As a result, we followed a more structured selection process, outlined in five steps.

*Step 1: Choosing Keywords*

We started by brainstorming keyword combinations that would help us find AI-integrated chatbots specifically in the healthcare domain. Without using terms such as "AI," "health," or "chatbot," the search results were flooded with unrelated apps such as fitness trackers, calorie counters, and heart rate monitors. The most relevant results came from combinations of "healthcare chatbot," "AI healthcare," and "AI health." Therefore, we used these keywords to cast a wide but focused net for relevant apps.

*Step 2: Initial App List*

Using a typical user approach, we searched apps on both the Google Play Store and Apple App Store, focusing on the most visible results. This was based on reports by Dogtiev, (2025) and MobileAction, (2025) which show that most users rely heavily on search and rarely scroll beyond the top few results. For example, 90% of Apple App Store users only view the top 10% of search results. To reflect this user behavior, we limited our selection to the first 20 apps shown for each keyword on each store. While each keyword produced over 100 results on both platforms (exact numbers are not available since platforms do not display total results), this method allowed us to compile an initial list of 120 apps, 60 from the Apple App Store and 60 from Google Play.

*Step 3: Apps on Both Platforms and Additional Reference Check*
We compared the lists side by side and highlighted apps available on both the Apple App Store and Google Play Store to ensure inclusion for both iOS and Android users. This resulted in 11 apps that matched our keywords and were included in the final selection.

To further validate our list, we cross-referenced our initial selections with a healthcare chatbot app list published in an article by Clark & Bailey, (2024) from the National Library of Medicine. This step helped confirm the relevance of five of our selected apps in the context of existing research.

*Step 4: Special Inclusion*
Lastly, we included a chatbot called "Health and Medicine," which is available within the ChatGPT platform. This chatbot provides specific medical information and, according to ChatGPT's usage data, has been accessed by over 200,000 users. Interestingly, ChatGPT itself appeared in all of our keyword searches, making this inclusion a natural part of the final selection.

*Exclusion Criteria and Selected Apps:*
We excluded apps based on the following criteria: (1) if they were only available on one platform; (2) if they did not have any healthcare-related features. For example, apps designed solely for entertainment or companionship, which still appeared in our search; (3) if they lacked AI chatbot functionality, even if they were healthcare-related, such as static information apps like "United Healthcare"; and (4) if they had fewer than 5K downloads on the Google Play Store. Although the Apple App Store provides ratings instead of download numbers, we did not exclude apps with low reviews as long as they had at least 5K downloads on the Google Play Store.

After applying all these criteria, a total of 12 from 120 apps were qualified. We randomly divided apps into two groups. We downloaded six apps on each device (iOS and Android). This approach ensured a balanced representation of applications across both platforms, and we aimed to explore if platforms have any impact on privacy practices (See Table 1). These selected applications include Youper, Wysa, Symptomate, Sensely, Jivi: Your AI Health Companion, Health Tracker: Healthily, ChatGPT (Health and Medicine by AIResearchPlus.com), Chatdok: Medical Chatbot, Chatbot AI - Healthcare (HealthGPT - Healthcare AI Chat), AI Therapist - Sintelly, AI Dermatologist: Skin Scanner, and ADA Health. We will identify these apps as App1…App12. *The identification of apps in this study is solely for methodological transparency and reflects neither endorsement nor censure.*

| Apps and Functions | Apple App Store | | Google Play Store |
|---|---|---|---|
| | Ratings | Star | Number of Downloads |
| App 1: Symptom checker | 11k | 4.8 | 10M+ |
| App 2: Therapy chatbot | 15k | 4.88 | 1M+ |
| App 3: Health self-care tracker | 393 | 4.2 | 1M+ |
| App 4: Mental health chatbot | 23k | 4.9 | 1M+ |
| App5: Therapy chatbot | 222 | 4.6 | 1M+ |
| App 6: Symptom checker | 131 | 4.6 | 500K+ |
| App 7: Health education chatbot | 200K+ | N/A | 200K+ |
| App 8: Skin condition analyzer | 4k | 4.6 | 100K+ |
| App 9: General Health assistant | 100< | 5 | 10K+ |
| App 10: Health advice chatbot | 100< | 5 | 10K+ |
| App 11: Health advice chatbot | zero rating | zero star | 5K+ |
| App 12: Health assistant | 137 | 4.6 | 5K+ |

**Table 1. AI Healthcare Apps on Apple and Google Play Stores**

**Evaluation Criteria**

We employed a structured privacy evaluation framework to assess the privacy practices of AI healthcare chatbot apps. We developed three evaluation checklists, Sign-Up Phase Evaluation Checklist, User Experience Phase Checklist, and Privacy Policy Analysis Checklist, to systematically analyze how these applications handle personal data. These checklists were designed with consideration of data protection regulations such as the GDPR (Articles 13–15), which is regarded as one of the strongest privacy protection laws in the world (Schünemann & and Windwehr, 2021) and CCPA which is the first comprehensive privacy regulation in the United States (US), and HIPAA which is a federal level health data protection regulation. In addition to legislation, we considered Fair Information Practice Principles (FIPPs) which are widely accepted standards for evaluating privacy (Fair Information Practice Principles (FIPPs), 1973). This process ensured that the checklists aligned with widely recognized privacy guidelines. To assess privacy policy transparency and compliance, we conducted a manual review of publicly available privacy policies of the chatbot apps.

Although there are several privacy frameworks such as LINDDUN, ISO 27701, and the NIST Privacy Framework, most of them are designed for internal organizational use and require access to system architecture, data flows, or backend documentation. Since we evaluated publicly available mobile apps from the app stores, we did not have access to these internal systems. Our evaluation approach focuses on what users can directly see and experience during app sign up, app use, and while reading the privacy policy. This makes it more practical for assessing real-world transparency from a user's perspective. The checklists we developed are not specific to AI and can be applied to a wide range of mobile apps. However, because we focused on apps that include AI chatbot features, we also examined what kinds of personal information the chatbot may ask users to provide during conversations. Future research could extend our framework by adding AI-specific checklist items. Our approach supports both legal alignment and user-centered evaluation by combining privacy regulations and principles such as GDPR, CCPA, HIPAA, and FIPPs.

To validate the consistency and clarity of the criteria, two researchers independently reviewed and pilot-tested the same chatbot app using the full set of checklists. Following this step, results were compared, discrepancies were resolved by consensus, and any ambiguous checklist items were refined based on discussion. To further assess the reliability of the coding framework, we conducted a formal inter-rater reliability check across 59 checklist items. The resulting Cohen's κ was 0.66 that showed moderate to substantial agreement. Then we finalized the codebook with definitions and supplementary notes to support consistent application across the remaining apps, which were then divided and evaluated independently by each researcher. For each app, we created a test user account and performed a structured walkthrough of its features. This included recording whether and how the app presented privacy notices, what personal or health information it requested, what permissions it prompted during use, and what privacy controls were available in settings. We interacted with each chatbot by asking common health questions and navigating the app a minimum 20 minutes to complete our checklist, follow up questions regarding health data, availability of data download or deletion within the app. We documented all observations with notes and screenshots for the records.

*Sign-up Phase Checklist*

The Sign-up Evaluation Checklist (see Table 2) assesses whether a privacy policy is presented during the sign-up process, it ensures that users are informed about data handling before they provide personal information. It also examines the types of Personally Identifiable Information (PII) requested during sign up, such as name, date of birth, email, and health-related data. Additionally, the checklist evaluates the permissions requested by the app, including access to the microphone, camera, location, and contact list. Lastly, it determines whether users can disable data sharing or manage permissions, allowing them greater control over their personal information before they start using the app.

| Privacy policy shown at sign-up? | Microphone access requested? |
|---|---|
| Requests PII during sign-up? | Camera access requested? |
| Full name requested? | Location access requested? |
| Date of birth requested? | Activity tracking toggle available? |
| Gender requested? | Data sharing opt-out at sign-up? |
| Email requested? | Contacts access requested? |
| Phone number requested? | Health data requested? (e.g., symptoms, meds) |
| Home address requested? | Tracks external activity? |
| PHI requested? (e.g., blood pressure, diabetes) | Manage permissions right after sign-up? |

**Table 2. Sign up Phase Checklist**

*User Experience Checklist*

We adopted an expert usability assessment where the two researchers interacted extensively with each chatbot. For each app, we created test accounts and performed a common set of actions: e.g., completing onboarding, asking the chatbot health-related questions, exploring settings, and attempting to use any privacy features. Each app was explored for specific durations (minimum 20 minutes) and scenarios (e.g. a symptom query scenario and a wellness advice scenario) to ensure depth.

The User Experience Evaluation Checklist examines the features available to users during app use, such as the ability to review, edit, or delete personal health records, this ensures users have control over their stored data. It also evaluates the types of PII requested during interactions with the chatbot, including name, date of birth, email address, and health-related information (see table 3) The checklist assesses the accessibility of privacy policies within the app, such as the number of clicks required to access them, as well as the availability of controls that allow users to manage app permissions, including location, microphone, and camera access. Lastly, it determines whether users have options to download or erase their personal data, which is essential for maintaining data autonomy and privacy.

| | |
|---|---|
| Location access requested? | Can users easily access and follow the privacy policy in the app? |
| If yes: Can users turn off/on location? | Can you download a copy of your personal data for your records? |
| Does the app ask for access to camera | Are there controls for mic, camera, or location settings? |
| Does the app ask for access to microphone | Can users edit or delete health data within the app? |
| Can we opt in/out while using app | Does the app provide device-level security (e.g., MFA)? |
| Can users access their health records in the app? | |

**Table 3. User Experience Checklist**

*Privacy Policy Evaluation Checklist*

The Privacy Policy Evaluation Checklist (see Table 4) assesses the accessibility of the privacy policy on the app's website to determine whether users can easily locate and review it. It also examines whether the privacy policy provides information regarding data collection, sharing practices, and retention periods, ensuring transparency in how user data is handled. Additionally, the checklist evaluates whether the privacy policy includes statements of compliance with key regulations such as GDPR, HIPAA, and CCPA, which establish legal standards for data protection. Lastly, it reviews the details provided about data security measures, including encryption protocols and breach notification policies, to assess how well the app protects users' sensitive information.

| | |
|---|---|
| Is the Privacy Policy available on the website? | Mentions encryption (at rest/in-transit)? |
| Does the Privacy Policy explain 3rd party data sharing? | Mentions access controls (e.g., MFA)? |
| Examples/agreements on data sharing? | Mentions other security (e.g., passwords)? |
| Mentions GDPR compliance? | Explains PHI handling? |
| Mentions HIPAA compliance? | Mentioning notifying users of PHI breaches? |
| Mentions CCPA compliance? | States exact data retention period? |
| States purpose for data collection? | Provide you methods to access, correct, or delete your data? |
| Mentions data minimization? | Informs about data rights (opt-out)? |

**Table 4. Privacy Policy Evaluation Checklist**

## RESULTS

We split our results into three sections: Sign-up Phase, User Evaluation Phase and Privacy Policy Evaluation Phase. Each section focuses on related results. Sign-up Phase results is indicating highlights of our major results during sign up process, User Experience Phase results summarize the main findings while using the apps, finally Privacy Policy Evaluation Phase results explain essential findings that we explored while analyzing AI healthcare chatbot privacy policies.

**Sign up Phase Results**

During the sign-up process, we found important differences in how the apps inform and protect users. Only half of the applications (50%) showed a privacy policy at the beginning, meaning the other half gave users no upfront information about how their data would be used. In terms of data collection, just 16% of the apps asked for personal details like date of birth, gender, or phone number during or shortly after sign-up. However, over half (58%) required an email address to create an account, and a quarter (25%) asked for personal or health-related information after sign-up, especially those focused on mental health.

When it came to permissions, very few apps requested access to a device's hardware at this stage. Only one app asked for camera access, two asked for the microphone, and none requested location data during sign up. About 25% of the apps asked for permission to track user activity across other apps or websites, while the rest avoided this type of tracking. Finally, only 16% of apps allowed users to say "no" to data sharing during sign-up. Another 41% gave some control over permissions later, but most users did not have full control from the start (see Figure 1).

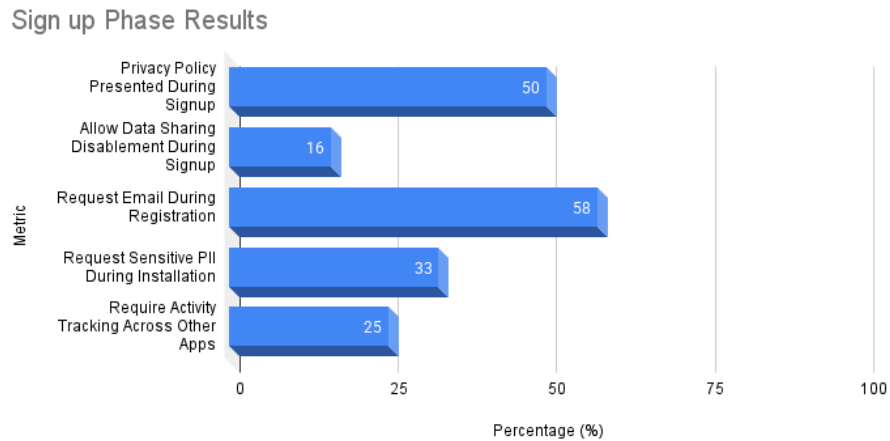

Figure 1. Sign up Phase Results by Percentage

**User Experience Phase Results**

When we began using the apps more as researchers, we saw additional requests for access and more privacy-related issues. Around 41% of the apps asked for microphone access, and 16% asked for camera access but these requests usually came later, not during sign up. Only 58% of the apps gave users an option to limit or turn off data collection while using the service. This suggests that many apps make it hard for users to change their privacy settings once they had started using them.

In terms of reviewing personal information, 67% of apps let users view their own health records that had been entered into the app. Half of the apps (50%) made it relatively easy to find their privacy policies, keeping them within three steps from the main screen. However, just 25% allowed users to download their health information. Security tools for protecting user accounts were rare, only 8% offered added features like two-step verification. We also found no meaningful differences in privacy features between the iOS and Android versions of these apps (see Figure 2).

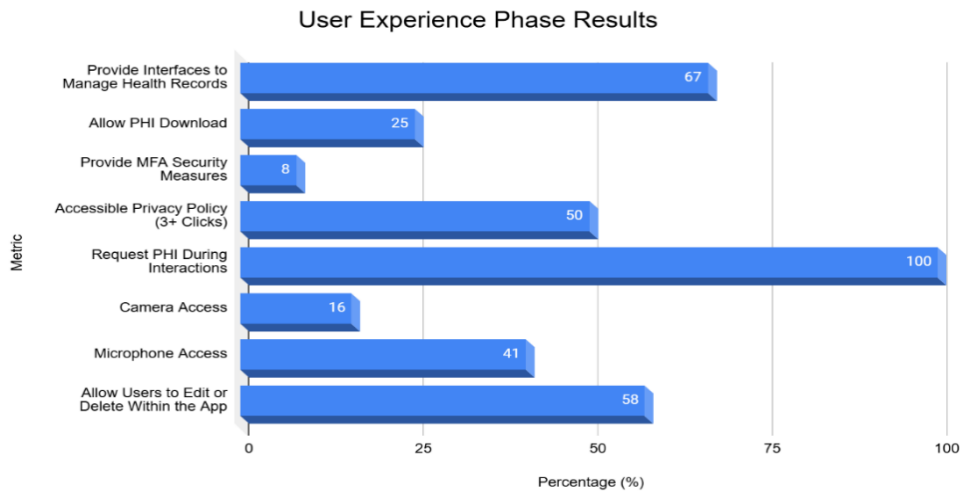

Figure 2. User Experience Phase Results by Percentage

**Privacy Policy Evaluation Phase Results**
Our review of the privacy policies found gaps in how these apps explain and follow existing laws. Only 8% of the apps said they follow HIPAA, the US health privacy law. While many of these tools may not be officially covered by HIPAA, it's still notable how few mentions it at all. Even more concerning, 42% of the apps did not mention GDPR, which is required for any service available to users in the European Union. Just over half (58%) referred to CCPA.

As for how the apps protect users' information, only 25% explained what security steps they use. About 20% mentioned encrypting data, and 33% described other safety measures like strong password rules or systems to prevent data loss. None of the apps included clear statements about what would happen if a data breach occurred. Most apps (91%) explained why they collect information, but only a third (33%) described how they handle sensitive health data specifically. In addition, only 41% of the apps told users how long their data would be stored, showing a lack of detail about how personal information is managed over time (see Figure 3).

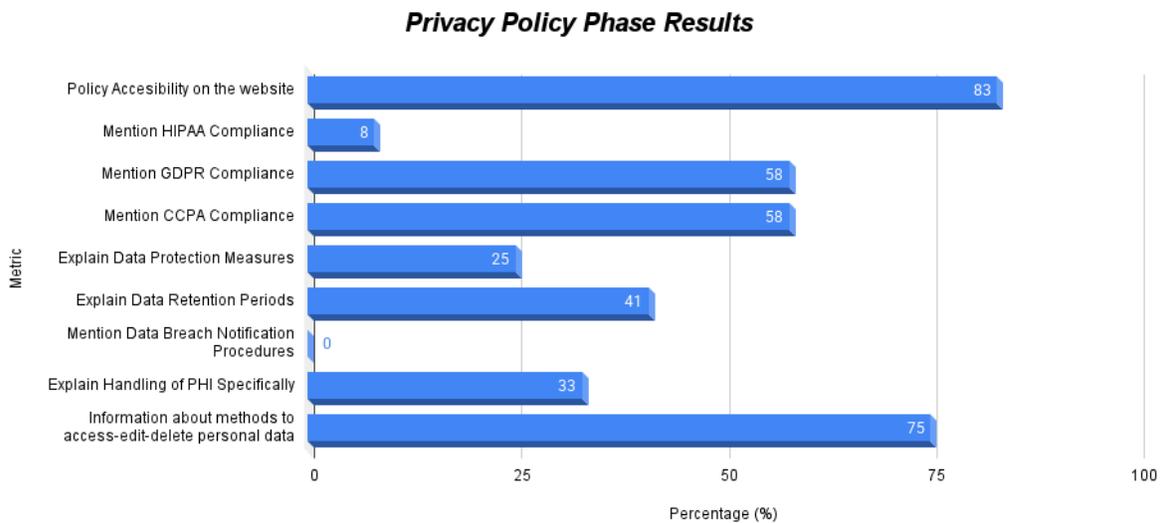

Figure 3. Privacy Policy Phase Results by Percentage

**DISCUSSION**
Our research shows significant privacy challenges regarding AI healthcare chatbot applications, in spite of their potential to enhance user experiences and improve healthcare interactions. When signing up, using the app or while interacting with the chatbots during the conversations, these applications collect extensive personal and health data. However, users have limited control over their own data. Only a small number of applications offer features such as disabling data sharing during sign up, downloading PHI, or editing and deleting personal data. This limited user agency suggests that, in many cases, the convenience and functionality of these chatbots come at the expense of individual privacy rights and informed consent. By not empowering users to easily manage their data, developers risk eroding the trust that is crucial for sustainable adoption of these technologies. Research by Zhang et al. (2024) demonstrated that users often fail to recognize privacy risks in AI-generated content, sometimes even preferring responses that include sensitive personal details over safer alternatives, they also found that people interacting with an AI agent ended up disclosing significantly more sensitive information. This shows that AI chatbots can encourage users to expose more personal information.

Privacy issues in AI healthcare chatbots raise broader ethical and legal questions. Many of these apps don't clearly fall under health privacy laws. If they are not run by official healthcare providers, they may not have to follow strict rules such as HIPAA in the U.S., even though they deal with very personal health information. Our analysis showed that their privacy policies often don't mention whether they follow health-related laws, which supports a common concern: current laws may not fully protect users when AI is involved in healthcare.

The analysis of privacy policies emphasizes inconsistencies in how transparent and accessible they are. While some applications make privacy policies accessible during sign-up or within a few steps while using the app, many of them fail to provide important details about data retention, security measures, or breach notification procedures. Such omissions leave users without sufficient information to make informed decisions about their data.

Moreover, we found that several apps fail to reference regulations such as GDPR, HIPAA, and CCPA, and even among those that do, the lack of specificity raises concerns about their compliance. These gaps in clarity and compliance not only undermine the integrity of the platforms but can also expose users to unrecognized risks, as personal data may be insufficiently protected or improperly managed. A more standardized approach to policy design, along with clearer adherence to relevant regulations, is essential for ensuring users understand the implications of their data sharing. These findings point out the urgent need for improvements in user-centric privacy practices and regulatory alignment. Apps should prioritize empowering users with greater control over their data, including intuitive interfaces to manage, edit, and retrieve personal information.

Privacy policies should adopt clear and standardized formats that comprehensively explain data handling practices, including collection, retention, security, and sharing. Such standardized approaches could leverage best practices drawn from established privacy engineering principles, presenting information in plain language and providing interactive controls. In addition, stronger enforcement of privacy regulations and mechanisms to certify compliance could enhance accountability and build user trust. By bridging the gap between current practices and the ideal standards set by laws and guidelines, stakeholders can promote a responsible ecosystem that respects user autonomy and fosters long-term engagement.

As we mentioned in the result section none of the apps explicitly mentioned data breach notifications in their privacy policies. The importance of data breach notifications for health-related apps comes from the Federal Trade Commissions' (FTC) Health Breach Notification Rule, which requires companies that collect sensitive health data (even those not directly covered by HIPAA) to inform users in case of unauthorized access or data breaches. Although most AI health chatbot apps we reviewed aren't considered HIPAA-covered entities, they still typically fall under the FTC's rule because they handle sensitive personal health information provided by consumers. Thus, the lack of clear statements about breach notifications is concerning, since users may not be aware of their rights or protections in case their personal health information is compromised.

Based on our observation, although our focus was not categorizing apps in terms of their niche fields, we observed that mental health apps selected for our study requested more detailed information about users during the chat while general health apps request less information about users; however we did not find a significant differences between general health apps and mental health apps in terms of privacy controls. We noticed that some apps have features allowing users to upload their laboratory results, doctor visit summaries, or any health-related documents. Even if the app doesn't require certain types of PII, it naturally gains access to that information once users upload these documents. This raises serious privacy concerns if the app doesn't have solid privacy practices in place. On top of that, users might share other personal information during conversations with AI chatbots. Even a simple question such as "How are you feeling today?" can lead users to reveal all sorts of personal details.

Our findings reflect long standing information science concerns around transparency and meaningful consent. From this perspective, privacy begins with data minimization and purpose limitation. Chatbots should collect only what is clinically essential and disclose how data will be used and for how long. Privacy policies must also address fairness and equity: breaches and misclassification errors can disproportionately harm under-represented communities whose data are already limited in training sets. Privacy enhancing technologies (PETs) such as differential privacy, federated learning, offer practical ways to reduce such risks while preserving utility.

Beyond technical solutions, Information Science also plays a role in promoting privacy awareness and digital literacy. If users do not understand how their data is handled or do not realize they have choices, it becomes harder to protect their privacy. The field can help by contributing to public education efforts, developing tools that make privacy settings easier to understand, and shaping ethical frameworks for AI in healthcare. Finally, as the legal landscape continues to evolve, information scientists can support efforts to ensure AI tools comply with existing regulations like GDPR, HIPAA, and CCPA, while also pushing for updates that reflect the unique challenges of conversational AI systems. Together, these contributions can help ensure that AI-powered health technologies respect user autonomy, foster trust, and are developed with strong ethical foundations.

**LIMITATIONS AND FUTURE RESEARCH**

One of the limitations of our study was the sample size of applications; we analyzed 12 apps which may not fully represent the variety of privacy practices in AI healthcare chatbots. Given the dynamic and rapidly evolving marketplace of digital health tools, a broader and more varied sample could uncover additional trends and highlight variations in compliance or transparency across different types of healthcare services. While our approach helped us to capture apps with substantial user bases, it may not reflect the practices of lesser- known or emerging chatbot apps that do not meet the same download thresholds. Our method focused mainly on apps that appeared in the top 20 search results in the Apple Store and Google Play Store, meaning we might have missed other relevant apps not prominently featured in search results. There's currently no good ranking system specifically for chatbot apps in these app stores, making systematic searching difficult. Although the Apple Store does have a medical category, it doesn't specifically focus on chatbot-related apps. Since we only reviewed a relatively small sample (12 apps), future research could include more apps or target specific areas, such as mental health chatbot apps, to provide a broader understanding of privacy risks. In addition, our evaluation relied on publicly available privacy policies and user-facing features.

Although these sources were helpful for our analyzes, it might not reveal behind-the-scenes data processing or undisclosed third-party partnerships.

Future research may address these limitations by analyzing a larger and more diverse pool of applications to ensure more representation of privacy practices. User surveys or interviews can help capture real-world experiences with chatbot apps. Including end-users in the study allows researchers to see if the policy language is clear, the interface is effective, and users feel confident managing their data. Expanding the geographic scope to include applications subject to different regulatory frameworks could provide a more comprehensive understanding of global privacy practices. Additionally, devel oping standardized frameworks for evaluating privacy practices in AI healthcare chatbots and exploring the relationship between improved privacy measures and user trust would also be valuable directions for future work. These frameworks could help guide developers, policymakers, and industry leaders in adopting best practices. Eventually, this could lead to a more harmonious balance between innovation, convenience, and the robust privacy protections users deserve, ultimately reinforcing confidence in the growing field of AI-driven health- care solutions.

It is important to note that all apps in our sample were accessed from the United States, and therefore we were not able to evaluate whether these apps adjust their privacy practices or user interfaces based on geographic region. Future research could explore whether the same apps behave differently when downloaded or accessed from jurisdictions with stricter data protection laws, such as the European Union. This would help assess whether stronger regulations translate into better user-facing transparency in practice.

**CONCLUSION**

In this study we study analyzed the privacy practices of 12 AI powered healthcare chatbot apps. As a result, we identified several critical gaps that might affect users' privacy, including inadequate transparency during the sign-up process, limited user control over personal data, and inconsistent compliance with essential regulations such as GDPR, HIPAA, and CCPA. We also found that the privacy policies of most chatbot applications lacked clearly stated cybersecurity measures or explicit procedures for managing data breaches, and this leaves users vulnerable and uninformed.

These issues demonstrate the significant ethical and practical challenges that arise as artificial intelligence becomes increasingly integrated into healthcare interactions. Information Science researchers and professionals have a crucial role in addressing these challenges by advocating for clearer data practices, improved transparency, and robust standards for user data protection. Information Science researchers can guide the responsible development of human-centered AI technologies by fostering collaboration among technology developers, policymakers, and end-users.

Future research can build upon these insights by exploring a broader and more diverse set of healthcare chatbot applications, investigating actual user experiences regarding privacy interactions, and developing standardized frameworks for privacy evaluations. Through such initiatives, the Information Science community can effectively advance the ethical deployment of AI in healthcare and other domains, and this helps ensure that these technologies not only drive innovation but also respect and protect user privacy and trust.

**GENERATIVE AI USE**

We employed generative AI, specifically ChatGPT, only for language clarity and grammar purposes. The authors assume all responsibility for the content of this submission.